\documentclass[prb,amsmath,amssymb,superscriptaddress,reprint,twocolumn]{revtex4}

\usepackage[utf8]{inputenc}

\usepackage{hyperref}
\usepackage{dcolumn}
\usepackage{bm}
\usepackage{hyperref}
\usepackage[dvipsnames]{xcolor}
\usepackage{graphicx}
\usepackage[capitalise]{cleveref}
\usepackage{textcomp}
\usepackage{pdfcomment}
\usepackage{bold-extra}

\DeclareUnicodeCharacter{2009}{\,}
\hypersetup{pdftoolbar=true,        
 pdfmenubar=true,        
 pdffitwindow=false,     
 pdfstartview={fitH},    
 pdftitle={},    
 pdfauthor={},     
 pdfsubject={DWsynthesis},   
 pdfcreator={},   
 pdfproducer={LaTeX}, 
 pdfkeywords={Version 1.0},
 colorlinks=true,       
 linkcolor=blue,          
 citecolor=blue,        
 filecolor=blue,      
 urlcolor=blue}

\makeatletter
\renewcommand{\@biblabel}[1]{#1. }
\renewcommand{\@dotsep}{500}
\renewcommand{\@pnumwidth}{0em}
\renewcommand{\l@figure}[2]{
\@dottedtocline{1}{1.5em}{2em}{Figure #1}{}\vspace{15pt}}

\usepackage[normalem]{ulem}

\begin{document}

\title{Highly-twisted states of light from a high quality factor photonic crystal ring}

\author{Xiyuan Lu}\email{xiyuan.lu@nist.gov}
\affiliation{Microsystems and Nanotechnology Division, Physical Measurement Laboratory, National Institute of Standards and Technology, Gaithersburg, Maryland 20899, USA}
\affiliation{Joint Quantum Institute, NIST/University of Maryland, College Park, Maryland 20742, USA}
\author{Mingkang Wang}
\affiliation{Microsystems and Nanotechnology Division, Physical Measurement Laboratory, National Institute of Standards and Technology, Gaithersburg, Maryland 20899, USA}
\affiliation{Department of
Chemistry and Biochemistry, University of Maryland, College Park, Maryland 20742, USA}
\author{Feng Zhou}
\affiliation{Microsystems and Nanotechnology Division, Physical Measurement Laboratory, National Institute of Standards and Technology, Gaithersburg, Maryland 20899, USA}
\affiliation{Joint Quantum Institute, NIST/University of Maryland, College Park, Maryland 20742, USA}
\author{Mikkel Heuck}
\affiliation{Department of Electrical and Photonics Engineering, Technical University of Denmark, Lyngby 2800 Kgs., Denmark}
\author{Wenqi Zhu}
\affiliation{Microsystems and Nanotechnology Division, Physical Measurement Laboratory, National Institute of Standards and Technology, Gaithersburg, Maryland 20899, USA}
\author{Vladimir A. Aksyuk}
\affiliation{Microsystems and Nanotechnology Division, Physical Measurement Laboratory, National Institute of Standards and Technology, Gaithersburg, Maryland 20899, USA}
\author{Dirk R. Englund}
\affiliation{Department of Electrical Engineering and Computer Science, Massachusetts Institute of Technology, Cambridge, Massachusetts 02139, USA}
\author{Kartik Srinivasan} \email{kartik.srinivasan@nist.gov}
\affiliation{Microsystems and Nanotechnology Division, Physical Measurement Laboratory, National Institute of Standards and Technology, Gaithersburg, Maryland 20899, USA}
\affiliation{Joint Quantum Institute, NIST/University of Maryland, College Park, Maryland 20742, USA}

\date{\today}

\begin{abstract}
      \noindent Twisted light with orbital angular momentum (OAM) has been extensively studied for applications in quantum and classical communications, microscopy, and optical micromanipulation. Ejecting the naturally high angular momentum whispering gallery modes (WGMs) of an optical microresonator through a grating-assisted mechanism, where the generated OAM number ($l$) is the difference of the angular momentum of the WGM and that of the grating, provides a scalable, chip-integrated solution for OAM generation. However, demonstrated OAM microresonators have exhibited a much lower quality factor ($Q$) than conventional WGM resonators (by $>100\times$), and an understanding of the ultimate limits on $Q$ has been lacking. This is crucial given the importance of $Q$ in enhancing light-matter interactions, such as single emitter coupling and parametric nonlinear processes, that underpin many important microresonator applications. Moreover, though high-OAM states are often desirable, the limits on what is achievable in a microresonator configuration are not well understood. Here, we provide new physical insight on these two longstanding questions, through understanding OAM from the perspective of mode coupling in a photonic crystal ring, and linking it to the commonly studied case of coherent backscattering between counter-propagating WGMs. In addition to demonstrating high-$Q$ ($10^5$ to $10^6$), high estimated OAM ejection efficiency (up to $90~\%$), and high-OAM number (up to $l$~=~60), our empirical model is supported by experiments and provides a quantitative explanation for the behavior of $Q$ and OAM ejection efficiency with $l$ for the first time. The state-of-the-art performance and new understanding of the physics of microresonator OAM generation will open new opportunities for realizing OAM applications using chip-integrated technologies.    
\end{abstract}

\maketitle
\noindent Light with orbital angular momentum (OAM)~\cite{shen2019optical,Erhard_NatRevPhys_2020}, previously known as helically phased light~\cite{allen1992orbital,coullet1989optical}, has been of long-standing interest. As an intrinsic property of photons, OAM with quantum number $l$ provides an additional dimension to encode information~\cite{winzer2014making}. This extra information capacity has been harnessed in holography~\cite{ren2019metasurface,ren2020complex,jack2009holographic}, multiplexed communications~\cite{Wang_NatPhoton_2012, hui2015multiplexed, xie2018integrated,Willner_APLPhoton_2021}, quantum entanglement~\cite{Nagali_PRL_2009,romero2012increasing,fickler2012quantum} and cryptography~\cite{mirhosseini2015high, Sit_Optica_2017}. After Allen \textit{et al.} pointed out that OAM is a natural property of all helically phased beams~\cite{allen1992orbital}, it has been routinely generated in free-space based on traditional helical beam generation methods~\cite{allen1992orbital,sueda2004laguerre,bauer2015observation}. Recently, thin film metasurfaces have been used as a single layer alternative to more traditional multi-level phase plates~\cite{Karimi_LSA_2014,bai2020high,ji2019dual}, and OAM light with $l$ up to 276 has been shown~\cite{bahari2021photonic}. In addition, spiral phase mirrors have been used to generate photons carrying OAM with $|l|$ $>$ 10,000, with its OAM ($\pm$ $l$) entangled with another photon's horizontal/vertical (H/V) polarization~\cite{fickler_PNAS_2016}.

On-chip OAM generation using integrated photonics~\cite{Yang_APLPhoton_2021} can advance more widespread use of OAM functionalities, and one major approach in this regard is through whispering gallery mode (WGM) microresonators~\cite{matsko_whispering_2005}. The WGMs in such resonators are bound modes that naturally support high angular momentum, and OAM-carrying states can be realized if a suitable means to eject such WGMs into free-space is incorporated, e.g., through a grating inscribed on the resonator~\cite{cai2012integrated}. For a WGM with azimuthal order $m$, a grating with $N$ periods around the resonator circumference will eject light carrying OAM with $l=m-N$.

The WGM approach is distinguished by the ability to simultaneously enhance light-matter interactions through the microresonator's high quality factor ($Q$) and small mode volume ($V$)~\cite{vahala_optical_2003}. This has been used, for example, in OAM semiconductor microlasers~\cite{miao_orbital_2016,Zhang_Science_2020} and in OAM single-photon sources based on the Purcell-enhanced emission of a single quantum emitter by the WGM~\cite{chen2021bright}. To maximize the microresonator's ability to enhance interactions while ejecting light into an OAM state, its high $Q$ should be retained even in the presence of the ejection grating, with the degradation in $Q$ relative to a conventional resonator (no grating) being exclusively due to the new coupling channel into the free-space OAM mode. This behavior should hold for a wide range of $l$, to fully enable the spatial multiplexing at the heart of OAM's potential in quantum and classical communications. However, existing demonstrations of OAM-generating microresonators have been limited to $Q\approx10^3$ so far~\cite{cai2012integrated,strain_fast_2014,chen2021bright}, and have focused on relatively low-$l$ OAM states. Moreover, quantitative understanding of the relationship between $Q$ and OAM ejection efficiency and $l$ has been lacking, and the full potential of such devices has remained unexplored.

Improving $Q$ in OAM-generating resonators has numerous implications. For example, in single quantum emitter systems, higher $Q$s would produce stronger Purcell enhancement to improve the indistinguishability and spontaneous emission coupling fraction of OAM single photons, with the further possibility of entering the non-perturbative strong coupling regime of cavity QED~\cite{lodahl_interfacing_2015}. A second example is spatiotemporal shaping of light~\cite{forbes_structured_2021}, where the ability to control both the spatial and temporal degrees of freedom of light is of both fundamental interest and can lead to new abilities for optical manipulation~\cite{dholakia_shaping_2011}. Recently, dynamic spatiotemporal control has been explored in the context of coherent addition of optical frequency comb components that carry different amounts of OAM~\cite{Zhao_NatCommun_2020}. Recent advances in frequency comb generation through nonlinear wave mixing in microresonators~\cite{kippenberg_dissipative_2018} suggest its potential in such research, but the limited $Q$s of OAM microresonators and the lack of understanding of these limits has prevented any serious investigation of such opportunities.

Here, we demonstrate chip-integrated, high-$Q$ ($10^5$ to $10^6$) microresonators that generate high-$l$ OAM states (up to $l=60$) with high estimated ejection efficiency (up to $90~\%$). We also provide a model that predicts the OAM ejection efficiency and microresonator's total dissipation rate and scaling with $l$. We do so by considering how OAM generation is one manifestation of grating-assisted coupling in a microresonator. In particular, we establish a connection between OAM ejection and mode-selective backscattering, known as selective mode splitting (SMS)~\cite{lu2014selective}, and show how measurements of SMS devices enable quantitative predictions of OAM behavior that are well-matched by experiments. Along with performance that dramatically exceeds previous studies in terms of $Q$ and accessible OAM states, our work provides a foundation for further development of OAM generation, particularly in the context of nonlinear and quantum light sources.   

\begin{figure*}[t!]
\centering\includegraphics[width=0.95\linewidth]{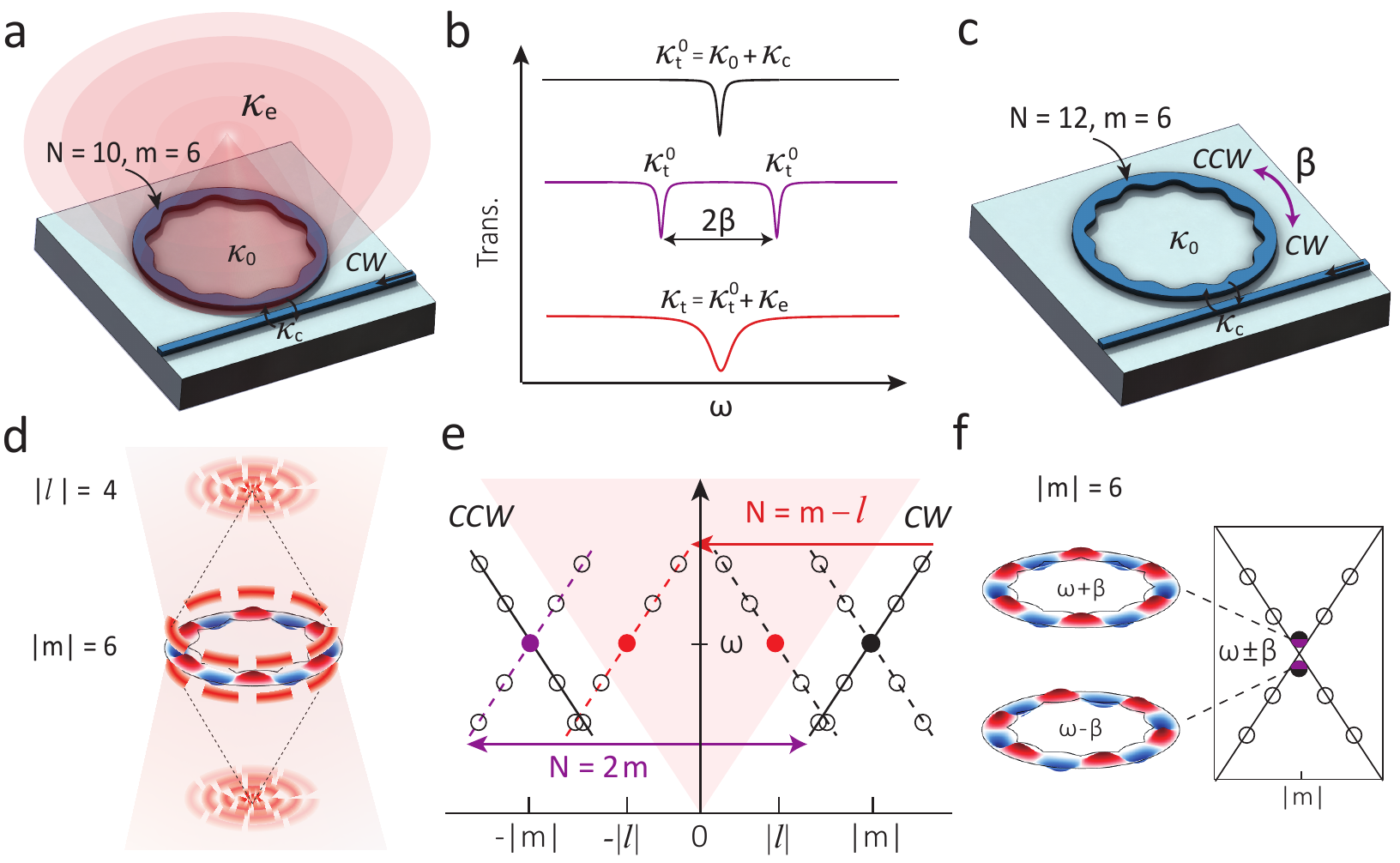}
\caption{Linking orbital angular momentum (OAM) emission with selective mode splitting (SMS) in a photonic crystal microring (PhCR). \textbf{a-c} Schematic of two PhCRs with $N = 10$ (a) and $N = 12$ (c), where $N$ is the number of modulation periods in a round trip. Their transmission spectra are illustrated in (b). Here, we focus on a whispering gallery mode (WGM) with an azimuthal mode number $m = 6$. Similar to a conventional (unmodulated) microring, this WGM has an intrinsic loss rate of $\kappa_0$ (from incoherent scattering and absorption) and is evanescently coupled by a waveguide with a coupling rate $\kappa_c$. The laser injected from the right side of the waveguide excites the clockwise WGM. In (a), the PhCR with $N = m + 4 = 10$ leads to OAM light emission, at an ejection rate $\kappa_e$, and carrying angular momentum of $l = -4$. In (b), the PhCR with $N = 12$ leads to a coupling of clockwise and counter-clockwise WGMs with $m  = \pm 6$, at a coupling rate $\beta$. \textbf{d} In the OAM device, the grating with $N = 10$ ejects the clockwise mode with $m = 6$ into a free-space OAM mode carrying a momentum of $l=-4$ at a rate of $\kappa_e$. The OAM emission leads to a broadening of the cavity linewidth ($\kappa_t = \kappa_t^0+\kappa_e$). When the WGM is a standing-wave with $m=\pm6$, or there is a reflection in the microring or chip facet (e.g., reflecting $m = 6$ to $m = -6$), the observed OAM mode has $l=\pm4$, which is manifested in its intensity profile exhibiting 4 pairs of anti-nodes. \textbf{e} Schematic band diagram for OAM and SMS. The purple double-arrow indicates the coupling between the clockwise and counter-clockwise traveling wave WGMs. The red single-arrow indicates the one-way ejection of light from the clockwise mode to the free space OAM mode. The OAM emission is symmetric to clockwise and counter-clockwise modes because the grating is static. \textbf{f} In the SMS device, the grating with $N = 12$ couples the clockwise and counter-clockwise modes with $m = \pm 6$ to each other without introducing excess loss (that is, $\kappa_t^0 = \kappa_0 +\kappa_c$), but introduces a mode splitting of $2\beta$ between the two renormalized standing-wave modes.}
\label{Fig1}
\end{figure*}

\medskip
\medskip
\noindent {\large\textbf{Results}}\\
\noindent \textbf{Principle Idea.} OAM ejection from a WGM is well-understood at a qualitative level, based on the basic angular momentum conservation criterion between the initial WGM with angular momentum $m$, the imprinted grating with $N$ periods along the ring circumference, and the resulting ejected OAM state with $l$=$m-N$~\cite{cai2012integrated}, as illustrated in Fig.~\ref{Fig1}(a, d). However, the key missing point is an understanding of the strength of the coupling from the WGM to the free-space OAM mode, which we quantify by a rate $\kappa_e$. This coupling leads to additional broadening of the total cavity linewidth, given by $\kappa_t = \kappa^0_t+\kappa_e$, where $\kappa^0_t$ includes the WGM intrinsic loss rate $\kappa_0$ and waveguide coupling rate $\kappa_c$, which is well-understood in conventional microrings. Such broadening is illustrated in Fig.~\ref{Fig1}(b). 
On the other hand, the interaction rate between two counter-propagating WGMs mediated by an imprinted grating, termed as selective mode splitting (SMS), is well-understood at a quantitative level~\cite{lu2014selective}. Here, we use a photonic crystal ring (PhCR) as an example, as shown in Fig.~\ref{Fig1}(c). The inside radius of the PhCR is modulated as $R_{in} = R^0_{in} + A\cos (N\phi)$, where $R^0_{in}$ is the average inside radius, $A$ is the modulation amplitude, $N$ is the number of periods of the grating, and $\phi$ is the azimuthal angle. Each WGM in the PhCR is characterized by an azimuthal mode number $m$, representing its angular momentum, that is, the number of electric field oscillations around the device perimeter within one round trip.  When $m=N/2$, the clockwise and counterclockwise WGMs are coupled by the photonic crystal grating. This coupling renormalizes two propagating modes into two standing-wave modes that see a narrower and a wider ring on average, and therefore have a smaller and larger resonance wavelength, or equivalently, a higher and lower center resonance frequency ($\omega_\pm = \omega_0 \pm \beta$), respectively, as illustrated in Fig.~\ref{Fig1}(f), where $\omega_{0}$ is the uncoupled (clockwise or counter-clockwise propagating) mode frequency. The coupling rate $\beta$ is simply given by $\beta = g A$, where $A$ is the modulation amplitude of the inside radius and $g=\partial\omega/\partial R_{in}$ at $R_{in} = R^0_{in}$, with $\omega$ the angular frequency of the WGM. We note that $g$ can be intuitively understood as the geometric dispersion with respect to the inside radius of an unmodulated ring. It is also equivalent to the per photon force (divided by $\hbar$) on the inside boundary of an unmodulated ring. Importantly, SMS WGMs remain high-$Q$~\cite{lu2014selective,lu2020universal,lu2022high} ($Q_t^{0} = \omega/\kappa_t^{0}$), with $\kappa_t^{0}$ remaining the same as a conventional microring, that is, $\kappa_t^{0} = \kappa_0+\kappa_c$, as illustrated in Fig.~\ref{Fig1}(b).

The contrast of the poor understanding of $\kappa_e$ in OAM with the clear understanding of $\beta$ in SMS is striking when we consider the similarity of these two systems, namely, that the number of periods in the grating ($N$) is the only difference in device geometry, with $N = m - l$ for the OAM light carrying $l$ momentum, and $N = 2 m$ for SMS. The geometries of the OAM and SMS devices are illustrated in Fig.~\ref{Fig1}(a,c), with their momentum-frequency diagrams shown in Fig.~\ref{Fig1}(e). The OAM mode is ejected from the device and cannot interact with the WGM mode after emission, as shown in Fig.~\ref{Fig1}(a), while clockwise and counter-clockwise WGMs can scatter back and forth, as shown in Fig.~\ref{Fig1}(c). In the band diagram shown in Fig.~\ref{Fig1}(e), the OAM emission is illustrated by a red arrow and the SMS coupling is illustrated by a purple double-ended arrow, assuming the waveguide initially couples light into the WGM in the clockwise direction. Figure~\ref{Fig1}(b) shows the expected transmission spectra of the control device (without modulation), the SMS case, and the OAM case. Compared to the control device, the SMS device shows a frequency splitting but no linewidth broadening, while the OAM device shows a linewidth broadening but no frequency splitting.

From coupled-mode equations for OAM and SMS (see the Supplementary Information for details), we propose a link between OAM and SMS given by:
\begin{equation}
\kappa_e = q_0\frac{2\beta}{\sqrt{F_t/(2\pi)}}  \cos(\theta). \label{Eq1}
\end{equation}
$\kappa_e$ and $\beta$ have the same units (both are rates), while all other parameters here are unitless. $q_0$ is a constant, and $F_t$ is the cavity mode finesse given by $F_t = Q_t/m = \omega/(m \kappa_t)$, where $Q_t$, $\omega$, and $\kappa_t$ are the total optical quality factor, cavity resonance angular frequency, and total cavity linewidth, respectively, of the corresponding WGM mode with an angular momentum of $m$ in the microring and an angular momentum of $l = m-N$ in the OAM emission. $\theta = (l/m) (\pi/2)$ represents the nominal twisted angle of the ejected OAM modes with respect to the vertical direction. Writing $\kappa_t$ in terms of its original value with no OAM emission ($\kappa^0_t$) and  OAM emission rate ($\kappa_e$), we get:
\begin{equation}
\kappa_t = \kappa^0_t + 2 q\sqrt{\kappa_t}. \label{Eq2}
\end{equation}
$\kappa^0_t$ includes the cavity intrinsic loss rate and waveguide-ring coupling rate, so that $\kappa^0_t = \kappa_0 + \kappa_c$. $q$ is related to $\kappa_e$ and $\kappa_t$ by $\kappa_e = 2 q \sqrt{\kappa_t} $, with $q = q_0 \beta \sqrt{m/\nu} \cos(\theta)$ (see Eq.~(\ref{Eq1})). Equation~(\ref{Eq2}) is a quadratic function of $\sqrt{\kappa_t}$, and its solution is given by:
\begin{equation}
\sqrt{\kappa_t} = q + \sqrt{q^2+\kappa^0_t}, \label{Eq3}
\end{equation}
where the other solution is negative and discarded.

From these simple equations, we can make a few initial observations. In the SMS case, where $l = -m$ ($N = 2m$), the cosine term vanishes, so that $q$ and $\kappa_e$ are zero. This is consistent with previous observations~\cite{lu2014selective,lu2022high} where $\kappa_t$ is barely affected by the grating modulation as long as $N=2m$. When $l = 0$, i.e., $N = m$, corresponding topologically to the $LG_{01}$ mode in the Laguerre-Gaussian basis of modes ($LG_{lp}$, where $l$ represents the angular momentum number and $p$ represents the radial momentum number), the cosine term is equal to one. In this case, when $\beta$ and $\kappa_e$ are small, the cavity linewidth asymptotically approaches that of the unmodulated microring ($\kappa_t \approx \kappa^0_t$). When $\kappa_e$ is large compared to $\kappa^0_t$, the OAM ejection channel is the dominant cavity loss channel ($\kappa_t \approx \kappa_e$). Finally, we posit that $\kappa_e\propto \cos(\theta)$, i.e, that the OAM ejection rate is linearly proportional to the momentum projected in the vertical direction after the grating's momentum is exerted on the WGM. This assumption requires experimental verification.
\begin{figure}[t!]
\centering\includegraphics[width=0.90\linewidth]{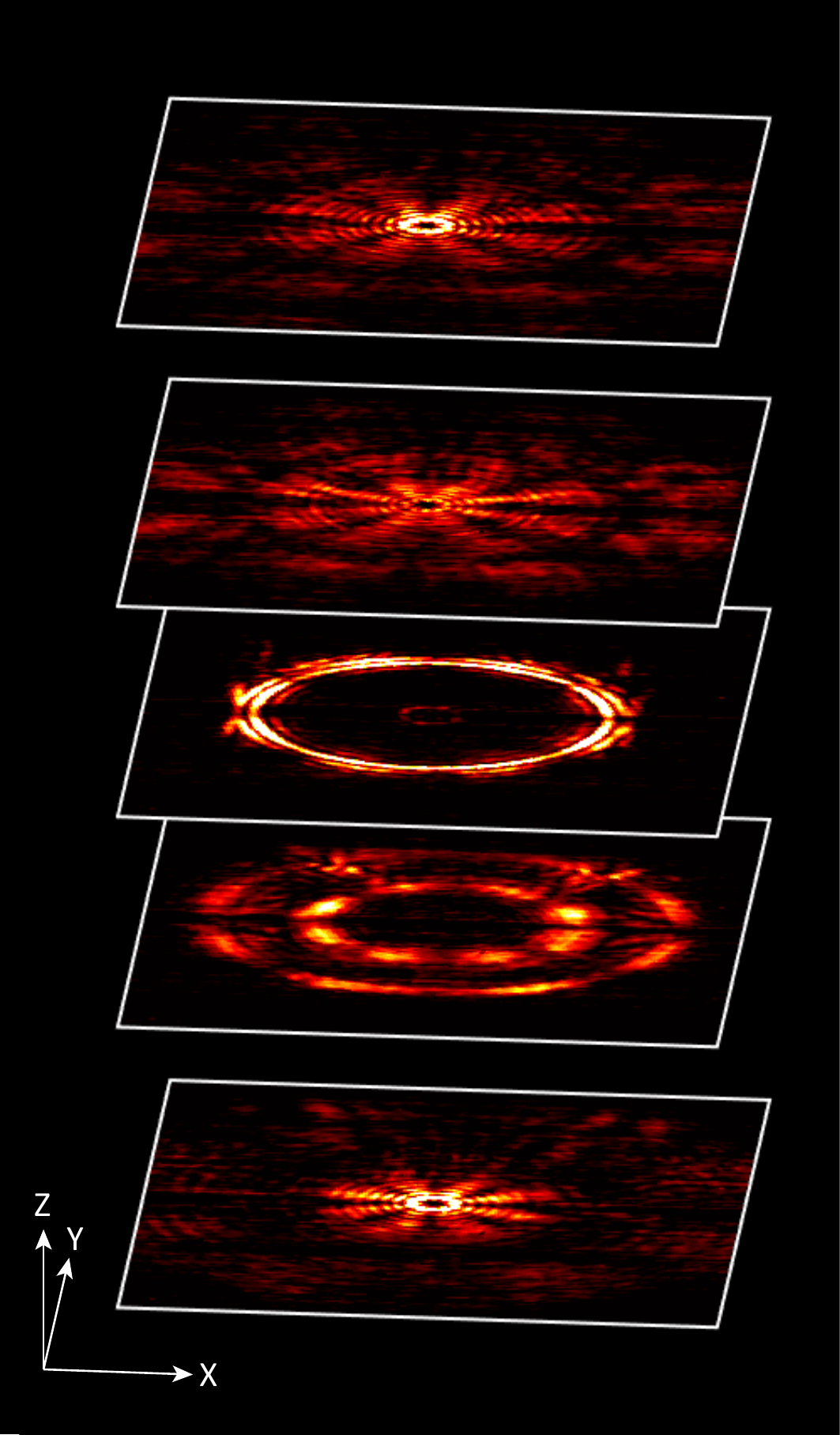}
\caption{OAM emission profile for $|l| = 4$. Experimentally measured mid-field and far-field infrared images of the $m = 165$ mode on a device with $N = 169$. These five images correspond to different heights of focal planes of the imaging system at approximately $\{$70, 30, 0, -30, -70$\}$~$\mu$m with respect to the microring. The square boxes corresponds to 80~$\mu$m$\times$100~$\mu$m on the x-y plane. The x-y-z arrows also serve as scale bars, whose lengths correspond to $\{$20, 50, 20$\}$~$\mu$m, respectively. }
\label{Fig2}
\end{figure}

\medskip
\noindent \textbf{Experimental examination from SMS to OAM.} We design and fabricate SMS and OAM devices in stoichiometric silicon nitride following the prescription of the previous section, with details provided in the Methods. Representative experimentally measured infrared images of the light ejected from one OAM device at various $z$ (vertical) planes are shown in Fig.~\ref{Fig2}. This device has $m$~=~165 and $N$~=~169, and the infrared images show OAM light with $|l|$~=~4. The interference patterns with 4$\times$2 nodes in both the mid-field and far-field are from the interference of OAM light with $l$~=~-4 and $l$~=~4 (the potential origin of this interference is discussed later). Going forward, we use such images to identify the $l$ number for each OAM state, while also considering $Q$ through transmission spectroscopy.

\begin{figure*}[htbp!]
\centering\includegraphics[width=0.95\linewidth]{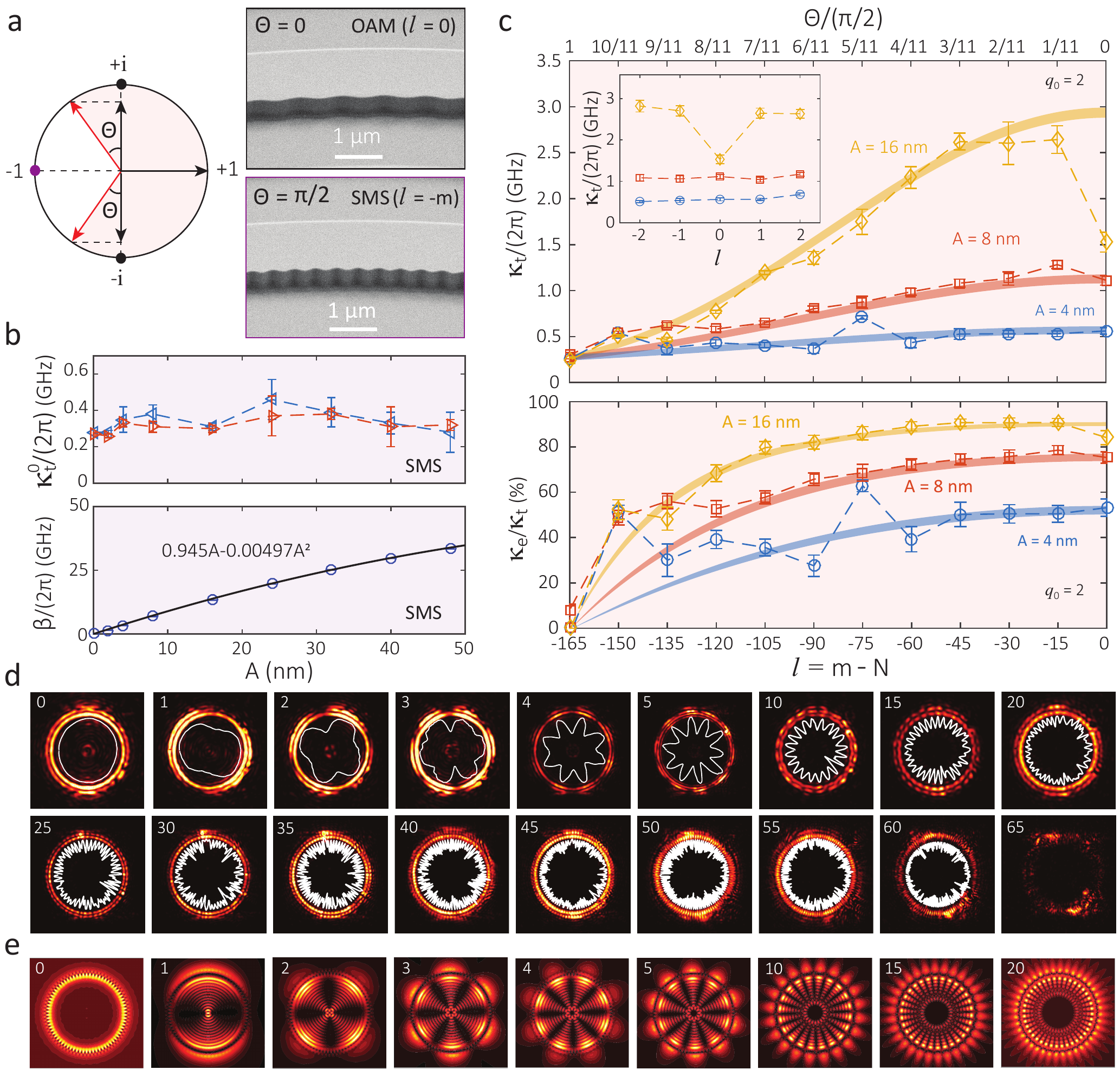}
\caption{Coherent emission of OAM light from high-Q whispering gallery modes. \textbf{a} (Left) Illustration of the angle $\theta$, which appears in Eq.~\ref{Eq1} in describing the OAM ejection rate $\kappa_e$, and (right) scanning electron microscope images of devices with $\theta=0$ (OAM) and $\theta=\pi/2$ (SMS), respectively. The top image shows approximately 5 periods of a PhCR for OAM ($N$=$m$=165), while the bottom image shows approximately 10 periods of a PhCR for SMS ($N$=2$m$=2$\times$165). Both devices have $A = 32$~nm. \textbf{b} In the SMS case, $\kappa_t^0$ of the two standing wave modes are not affected by increasing $A$, as shown in the top panel. The error bars are 95~\% confidence intervals of the nonlinear fits to the cavity transmission data (see Supplementary Fig.~S3 for the details). The mode splitting ($\beta$) is nearly linearly dependent on $A$ for the targeted WGM mode, whose experimental data and fit are shown in the bottom panel. The error bars are approximately 0.3~GHz, and are within the data symbols. \textbf{c} Plot of the OAM cavity linewidth ($\kappa_t$) and estimated OAM ejection efficiency ($\kappa_e/\kappa_t$) in the top and bottom graphs, respectively, as a function of $l = m - N$ on the bottom x-axis and $\theta/(\pi/2) = |l/m|$ on the top x-axis. We investigate three sets of devices having different modulation amplitudes of $A$ = \{4, 8, 16\}~nm, with $N$ varying from $2m$ to $m$ (i.e., transiting from the SMS regime to the OAM regime). We focus on a specific WGM at $\omega_0$ with mode number $m_0$ in each set of devices. The error bars represent the 95~\% confidence intervals from nonlinear least squares fits to the transmission data. The translucent shaded curves are from theoretical estimates predicted by Eqs.~(\ref{Eq1})-(\ref{Eq3}) with $q_0 = 2$, with all other parameters taken from measurements of OAM and SMS devices. The width of each shaded curve originates from the combined range of six fitted $\kappa^0_t$ values from three SMS devices at $l$~=~-165. The top inset shows a region of discrepancy between the experiments and the model, near $l$~=~0. The coupling waveguide in use has a nominal width of 750~nm and a microring-waveguide gap of 500~nm. \textbf{d} Infrared images near the surface of the OAM microrings for different $|l|$ values, with each displaying $2|l|$ anti-nodes. The angular intensity distribution is re-scaled and plotted in white to guide viewing. \textbf{e} Predicted patterns from three-dimensional finite-difference time-domain simulations using dipole excitation. While the OAM states are the same as in the experiments, a smaller device size (radius of 12~$\mu$m) and symmetric air cladding are used to keep the simulation size tractable.}
\label{Fig3}
\end{figure*}

We next consider the close connection between SMS and OAM devices, with representative devices shown in Fig.~\ref{Fig3}(a). The length of a modulation period, given by $2 \pi R/N$, is twice as long in this OAM device ($N = m$, i.e., $l = 0$) as in the SMS device ($N = 2m$), but all other parameters are kept the same. We fabricate a series of devices for SMS and OAM, varying $N$ while keeping the device geometry otherwise fixed. By studying modes of the same azimuthal order $m$ and similar resonance frequency $\omega$, we endeavor to limit the impact of any systematic variation in intrinsic and coupling $Q$ (e.g., with frequency, ring width, thickness, refractive index, etc), enabling us to focus on how $\kappa_t$ and $\kappa_e$ vary with $l=m-N$.

The SMS results are summarized in Fig.~\ref{Fig3}(b), and are consistent with previous reports~\cite{lu2014selective, lu2020universal,lu2022high}: the total cavity linewidths ($\kappa_{t}^0$) see no change to within measurement uncertainty when $A$ increases, and the mode splitting (2$\beta$) is essentially linearly dependent on $A$, when the splitting is $>$10$\times$ smaller than the free spectral range (approximately 1~THz in these devices). The error bars represent 95~$\%$ confidence intervals from nonlinear least squares fits to the SMS transmission data (see Supplementary). The measured $\kappa^0_t/(2\pi)\approx0.3$~GHz corresponds to a $Q^0_t\approx6.4\times$10$^5$ at 1560~nm. Using $\kappa_t^0$ and $\beta$ from SMS, we can predict the total OAM cavity linewidth ($\kappa_t$) and OAM ejection efficiency ($\kappa_e/\kappa_t$) through Eqs.~(\ref{Eq2})-(\ref{Eq3}), with only one free parameter $q_0$.

In the top panel of Fig.~\ref{Fig3}(c) we plot the measured $\kappa_t$ for a series of OAM devices, where $N$ has been varied so that $l$ ranges between -165 and 0, and for three different values of $A$. We find that this experimental behavior agrees well with our model using the measured SMS values and $q_0=2$, as shown by the different color solid curves in Fig.~\ref{Fig3}(c). The width of the curves represents the uncertainty in the predictions due to the uncertainties of $\kappa_t^0$ that come from nonlinear least squares fits to the SMS transmission data (see Methods). We note that the predictions deviate from experiments near $l=0$ for large $A$, with the inset zooming in on this behavior with adjacent $l$ from -2 to +2. This low-radiation-loss mode only happens at $l=0$, which has been used in integrated microrings for single-mode lasing~\cite{arbabi_grating_2015}, and its physics is related to a bound state in the continuum phenomenon ~\cite{hsu2016bound,yulaev2021exceptional} induced by the photonic crystal structure.

The bottom panel of Fig.~\ref{Fig3}(c) shows the estimated extraction efficiency $\kappa_e/\kappa_t$ as a function of $l$, where $\kappa_e$ is experimentally determined from the measured $\kappa_t$ (from the OAM devices) and the measured $\kappa_t^0$ from the SMS devices. The experimental data is again matched well by the model, particularly for larger values of $A$, where the model results are shown as solid curves whose widths are determined by the aforementioned uncertainties in the experimental SMS data. Importantly, the model contains no free parameter other than measured from experiments, except $q_0$~=~2, which represents the upward and downward OAM emission paths. Between the two panels of Fig.~\ref{Fig3}(c), we see the basic trend that the estimated OAM ejection efficiency and total cavity linewidth both increase in moving from $l=-165$ to $l=0$. The OAM ejection efficiency and total cavity linewidth also scale with modulation amplitude $A$ as expected, with the level of agreement between theory and experiments improving with increasing $A$. The estimated ejection efficiency reaches $\kappa_e/\kappa_t$~$=$~(80$\pm$3)~$\%$ at $l = -105$ and $A =$~16~nm, with $\kappa_t/(2\pi)$ of (1.19$\pm$0.02) GHz and thus $Q_t$ of (1.62$\pm$0.02)$\times$10$^5$. This efficiency is further increased to $\kappa_e/\kappa_t$~$=$~(90$\pm$1)~$\%$ at $l = -15$, with a broadening of $\kappa_t$ to (2.6$\pm$0.2)~GHz governed by Equation~(\ref{Eq3}).

We also perform imaging of the OAM microring modes to confirm their spatial behavior as a function of $l$. As noted earlier, Fig.~\ref{Fig2} shows the results for a microring with $m$=165 and $N$=169. Rather than a pure $l=-4$ state, the images are consistent with the emission containing both $l=-4$ and $l=4$ contributions, resulting in 4$\times$2 antinodes in the measured distribution. Similar behavior has been observed in other OAM microcavity works~\cite{chen2021bright}, where it was attributed to ejection of light from a standing wave cavity mode. In our case, the ejection of both CW and CCW light could be due to surface roughness or waveguide facet reflection at the edge of the chip. The back-coupling rate of this reflection seems to be smaller than the total linewidth (unlike the SMS case), so a clear splitting of resonance is not observed in general. Next, Fig.~\ref{Fig3}(d) displays the imaged OAM microrings fields near the surface of the cavities for a variety of OAM states with increasing $|l|$, as determined by analyzing the images and counting the number of anti-nodes. OAM states from $|l|=0$ to $|l|=60$ are clearly observed; the observation of even higher-order OAM is likely limited by the numerical aperture of our imaging system. We note that in these measurements, devices with $|l| = $ 1 to 3 had an additional SMS modulation imprinted on the device pattern to ensure standing wave modes for better interference visibility; this method is discussed further in the next section. A comparison of devices with and without SMS is analyzed in Supplementary.

We compare our results against finite-difference time-domain simulations, with the simulation methods outlined in the Supplementary Information. Dipole excitation is used to excite standing-wave WGMs to have a beating pattern in the intensity for OAM. Figure~\ref{Fig3}(e) shows that the simulation results qualitatively agree with the observed patterns. Plotted here is the Poynting vector projected on the vertical direction, that is, $\mathbf{S}_z = (\mathbf{E} \times \mathbf{H}) \cdot \mathbf{\hat{z}}$, in the mid-field above the surface of the microring. The Supplementary Information provides further simulations of emitted OAM for both standing-wave and traveling-wave WGMs. 

Finally, we emphasize that the observed $Q$s, in addition to following the predicted trends based on the SMS devices and Eqs.~(\ref{Eq1})-(\ref{Eq3}), are more than two orders of magnitude higher than those demonstrated in previous OAM generators based on microring resonators~\cite{cai2012integrated,chen2021bright}, while simultaneously exhibiting a high estimated ejection efficiency. For example, the $|l|$~=~60 mode has $Q_t\approx5\times10^5$ and an estimated ejection efficiency of 40~$\%$ for $A$~=~4~nm and $Q_t\approx2\times10^5$ and an estimated ejection efficiency of 65~$\%$ for $A$~=~8~nm. Such high-$Q$s are particularly promising for enhancing light-matter interactions, for example, to create Purcell-enhanced quantum light with OAM from a quantum emitter~\cite{chen2021bright}, to realize coherent spin-photon interfaces~\cite{duan2021avertical}, or to mediate nonlinear wave mixing interactions such as Kerr comb generation and entangled-photon pair generation with the output fields encoded in OAM states~\cite{strekalov_nonlinear_2016}. 

\begin{figure*}[t!]
\centering\includegraphics[width=0.95\linewidth]{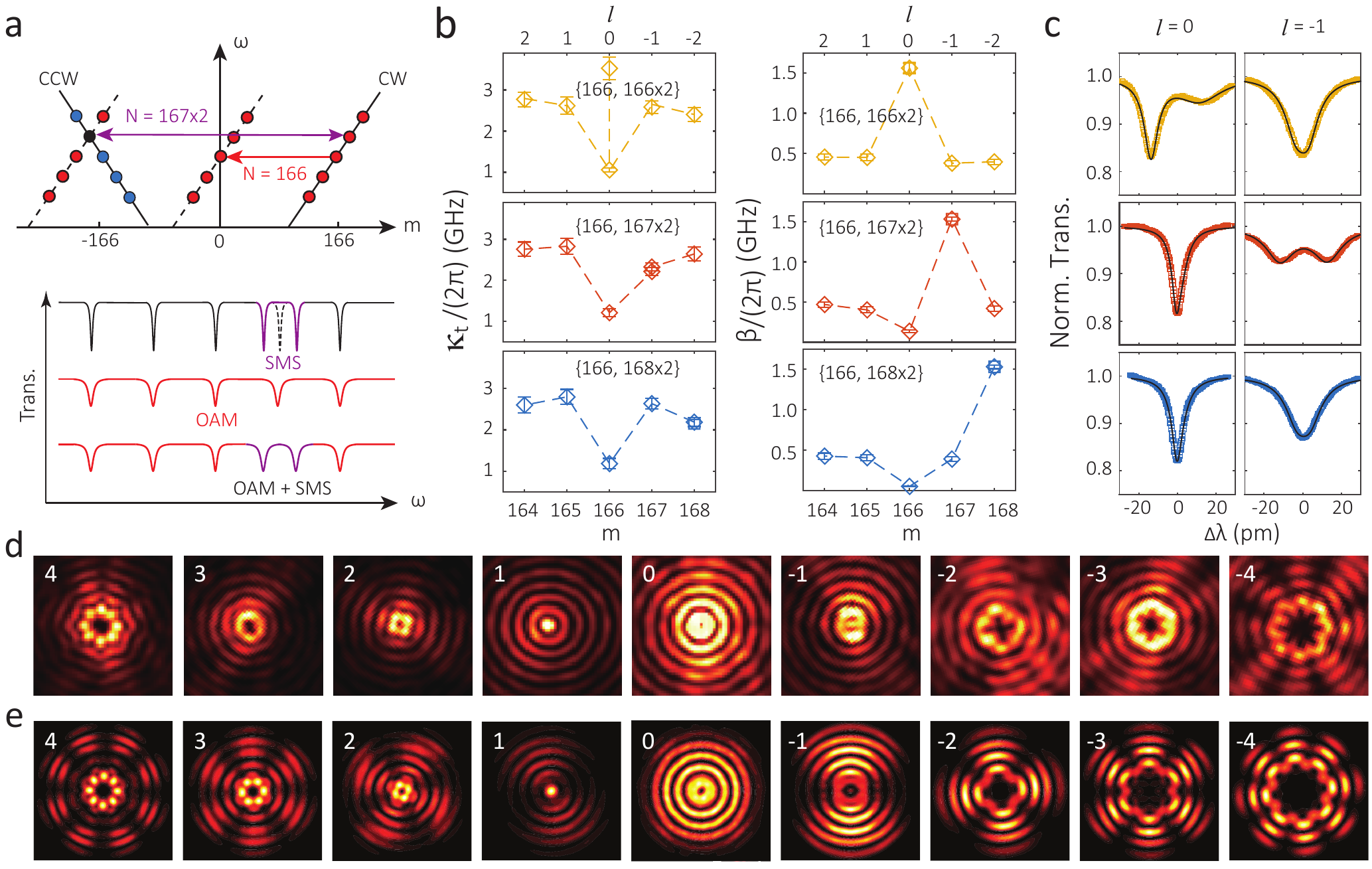}
\caption{Coherent implementation of OAM and SMS simultaneously. \textbf{a} Schematic band diagram illustration (top) and expected transmission spectrum (bottom) when selective mode splitting (SMS) and OAM are coherently implemented together. \textbf{b} Here, we have three devices with the same OAM modulation of $N =$ 166 and $A =$ 16~nm but different SMS modulation of $N = \{166, 167, 168\}\times2$ and $A$ = 2~nm in yellow, red, and blue, respectively. The left column displays the total dissipation rate from the cavity ($\kappa_t$) while the right column displays the backscattering rate ($\beta$), both as a function of azimuthal mode order ($m$) and OAM state ($l$), with both obtained from nonlinear fits to transmission spectra for each mode (the error bars represent 95~\% confidence intervals of the nonlinear fits). Two data points are displaced in $\kappa_t$ figures for the SMS modes, though overlapped within the error bars in the case of $m = 167\times2$ and $m = 168\times2$. \textbf{c} Normalized transmission spectra for the $l = 0$ and $l = -1$ modes (color) and their fitting curves (black) for each case, from which $\kappa_t$ and $\beta$ are extracted. \textbf{d} Images of the emitted light from the different $l$ orders take  at a the focus plane (70$\pm$10)~$\mu$m from the device. Different OAM states can be recognized by the distinctive `ripple' patterns generated. Here we show $l$ from 4 to -4 as examples. \textbf{e} Predicted far-field patterns from three-dimensional finite-difference time-domain simulations. Here $|\mathbf{E}|^2$ projected to the upper hemisphere with a radius of 1 meter is plotted, see the Supplementary Information for details. While the OAM states are the same as in the experiments, a smaller device size (radius of 5~$\mu$m) and symmetric air cladding are used to keep the simulation size tractable.}
\label{Fig4}
\end{figure*}

\medskip
\noindent \textbf{Combining SMS and OAM coherently.} So far we have been using a single-period grating for either SMS or OAM. Since both scattering processes are coherent, it is possible to combine them. For example, previous work has shown that combining multiple SMS periods through a multi-period grating (i.e. by simply adding up modulation with different $N$s) is practical and retains high cavity quality factors~\cite{lu2020universal}. Here we use a dual-period grating to implement SMS and OAM together. For comparison, we study three cases with a fixed number of modulation periods for OAM at $N = 166$ and a varying number of modulation periods for SMS at $N = 2\times\{166, 167, 168\}$. In the band diagram displayed in the top panel in Fig.~\ref{Fig4}(a), we illustrate the case in which the $m = 166$ mode is ejected to an $l = 0$ OAM state and the $m = \pm167$ modes are coupled via SMS. The resulting cavity transmission is illustrated in the bottom panel, where SMS splits the $m = 167$ mode (in purple) without affecting linewidths and OAM broadens all the cavity linewidths (in red). Having both SMS and OAM should result in a coherent summation of both effects.

We examine the implementation of coherent OAM and SMS in three fabricated devices, as shown in Fig.~\ref{Fig4}(b, c), where in contrast to the previous section, here we do not focus on a single azimuthal order mode, but instead examine a series of adjacent azimuthal order modes. Figure~\ref{Fig4}(c) shows representative transmission spectra (for $l$~=~0 and $l$~=~-1, or equivalently $m$~=~166 and $m$~=~167). Figure~\ref{Fig4}(b) shows the extracted loaded cavity linewidths ($\kappa_t$) created by OAM in the left column, and the right column shows the mode splitting ($\beta$) created by SMS. The overall behavior we observe is consistent with expectation for coherent superposition of the OAM and SMS effects. The mode splittings is largest for the azimuthal mode targeted by the SMS modulation, while the OAM modulation is set to eject the l=0 mode, and consistently shows a reduction in dissipation as observed in the previous section.

With or without SMS, our OAM devices always show standing-wave patterns in images taken both at the top surface of the microring, as shown in the previous section by Fig.~\ref{Fig3}(d), and in the far-field, as we show in Fig.~\ref{Fig4}(d). These standing-wave images resemble previous reports~\cite{chen2021bright,duan2021avertical}, and is not an issue in many quantum systems, as the emitted light is intrinsically in both clockwise and counter-clockwise directions~\cite{chen2021bright}. As noted earlier, their precise origin in our system requires further investigation. That being said, we find that the measured far-field images are in good agreement with the results of finite-difference time-domain simulations that incorporate a standing wave mode pattern, as shown in Fig.~\ref{Fig4}(e). 

Importantly, our results indicate that OAM emission does not have to lead to a mode splitting or a considerably broadened linewidth~\cite{cai2012integrated,strain_fast_2014}, while the purity of OAM emission and its impact on the OAM efficiency (our estimate given by $\kappa_{e}/\kappa_{t}$ is an upper bound) require further investigation. For example, the popular square grating is effectively a composition of multiple frequency components, while only the fundamental frequency grating (as we employ with a sinusoidal modulation) is essential for OAM. The potential role that such multi-frequency components play on excess loss and backscattering is still an open question, and to this end, our approach from SMS to OAM can be extended to these structures to perform a quantitative evaluation.

\medskip 
\medskip 
\noindent {\large \textbf{Discussion}}\\
\noindent We have demonstrated high-$Q$ optical microcavities with controllable and efficient OAM ejection. By linking OAM ejection to the closely related effect of selective mode splitting (SMS) due to backscattering in a microresonator, we present a predictive model for the OAM cavity linewidth and ejection efficiency. We showcase twisted light with $l$ from 0 to 60 and $p = 0$ (i.e., fundamental in the transverse direction), and it should be straightforward to extend to larger $l$ and $p$. Our results are of use to many photonics applications, including OAM multiplexing and entanglement for classical or quantum photonics applications. Future scientific understanding includes the origin of the coefficient $q_0$ that relates the OAM ejection rate to the cavity finesse, azimuthal mode number, OAM state, and backscattering rate for the analogous SMS device. Additional studies to undertake include the effects of the light cone(s) defined by the cladding and substrate layers on $\kappa_e$ and the origin of the apparent $+l$ and $-l$ superposition in the ejected light. Further important engineering tasks include using metamaterial structures with high numerical aperture to collect highly-twisted OAM light, collecting/multiplexing OAM light into the optical fibers, and using OAM states to control/manipulate atomic states on top of the photonic chip.

\medskip
\noindent {\large \textbf{Data availability}}\\
\noindent The data that supports the plots within this paper and other findings of this study are available from the corresponding authors upon reasonable request.

\bibliographystyle{naturemag}

\renewcommand{\figurename}{SI Fig.}
\renewcommand{\tablename}{SI Tab.}
\setcounter{figure}{0}

\medskip
\medskip
\noindent {\large \textbf{Methods}}\\
\noindent \textbf{Fabrication method}. The stoichiometric ${\rm Si_3N_4}$ layer is grown by low-pressure chemical vapor deposition with a nominal thickness of 500~nm on a ${\rm SiO_2}$ layer approximately 3~$\mu$m thick and grown via thermal wet oxidation of a 100~mm diameter Si wafer. The ${\rm Si_3N_4}$ layer thicknesses, as well as its wavelength-dependent refractive index, was confirmed using spectroscopic ellipsometry, with the index fitted to an extended Sellmeier model. A layer of positive-tone resist (approximately 650~nm in thickness) is spun on top of the ${\rm Si_3N_4}$ layer and exposed by a 100 keV electron-beam lithography system. The device layout is prepared using the Nanolithography Toolbox~\cite{Balram_JResNIST_2016}, a free software package developed by the Center for Nanoscale Science and Technology at the National Institute of Standards and Technology. The pattern in use has a resolution of 1/8~nm for the in-plane grids, and has an angular resolution of $\pi/max(N)/12$ for the inside modulation, where $max(N)$ corresponds to the largest number of cell numbers (i.e., smallest period lengths) in use. During the lithography, the minimal grids are further increased to 2~nm due to the shot pitch limitation of the electron beam system in use for a 500~pA electron current. We can observe selective mode splitting down to a nominal modulation amplitude $A$=1/8~nm using this method while maintaining a nearly linear dependence of mode splitting on amplitude. This observation is quite surprising considering the fracturing of the 2~nm shot pitch of the electron beam and its proximity effects, and it requires further study to clarify the underlying mechanism in terms of the actual pattern geometry relative to the designed one. Once the exposed pattern was developed, it was transferred to the ${\rm Si_3N_4}$ using a ${\rm CHF_3/O_2}$ chemistry by reactive ion etching, with a rate of approximately 30~nm per minute. The remnant resist and deposited polymer during the etching process are chemically cleaned. An SiO$_2$ lift-off process is performed so that the microrings have a top air cladding while the input/output edge-coupler waveguides have a top ${\rm SiO_2}$ cladding. Such top and bottom ${\rm SiO_2}$ claddings create more symmetric modes for coupling to optical fibers, reducing the fiber-chip facet coupling loss to 2~dB to 3~dB per facet. The oxide lift-off process is based on photolithography, plasma-enhanced chemical vapor deposition of SiO$_2$ with an inductively-coupled plasma source, and the chemical removal of the photoresist. After the lift-off process, the chips are diced and polished, and annealed at $\approx$~1000~${\rm ^{\circ} C}$ in a ${\rm N_2}$ environment for 4 hours.

\medskip
\medskip
\noindent {\large \textbf{Acknowledgements}}\\
\noindent The authors acknowledge Jin Liu, Haitan Xu, Zhimin Shi, and Amit Agrawal for helpful discussions. This work is supported by DARPA SAVaNT the NIST-on-a-chip programs, and partly sponsored by the Army Research Office under Cooperative Agreement Number W911NF-21-2-0106. M.W. is supported under the cooperative research agreement award 70NANB10H193, through the University of Maryland, College Park. M. H. acknowledges support from the Villum Foundation (QNET-NODES grant no. 37417).


\newpage
\onecolumngrid
\noindent{\large \textbf{~~~~~~~~~~~~~~~~~~~~~~~~~~~~~~~~~~~~Supplementary Information}}\\

\section{Coupled-mode Equations}
\noindent In this section, we give coupled-mode equations linking OAM and SMS, leading to Equation (1) of the main text. First, the coupled mode equations for the clockwise ($\circlearrowright$) and counter-clockwise ($\circlearrowleft$) waves in the resonator can be given by the following:
\begin{eqnarray}
\frac{d\tilde{A}_{\circlearrowright}}{dt} &=& (i\Delta\omega - \kappa_\text{0}/2)~\tilde{A}_{\circlearrowright} + i \beta ~\tilde{A}_{\circlearrowleft} + i \sqrt{\kappa_\text{c}} \tilde{S}_{\circlearrowright},  \label{EqS1} \\
\frac{d\tilde{A}_{\circlearrowleft}}{dt} &=& (i\Delta\omega - \kappa_\text{0}/2)~\tilde{A}_{\circlearrowleft} -  i \beta^* ~\tilde{A}_{\circlearrowright} + i \sqrt{\kappa_\text{c}} \tilde{S}_{\circlearrowleft}. \label{EqS2}
\end{eqnarray}
The field amplitudes are normalized so that $U_{\circlearrowright} = |\tilde{A}_{\circlearrowright}|^2$ and $U_{\circlearrowleft} = |\tilde{A}_{\circlearrowleft}|^2$ represent the intracavity optical energy. The cavity detuning is given by $\Delta \omega = \omega - \omega_0$, where $\omega$ is the angular frequency and $\omega_0$ is the central frequency of the cavity resonance. Only linear interaction and coupling terms are considered, and all nonlinear interaction terms are not included. We assume these two propagating waves have identical intrinsic loss rate ($\kappa_\text{0}$) and microring-waveguide coupling rate ($\kappa_\text{c}$). In the selective mode splitting (SMS) case, the backscattering induced by the photonic crystal structures ($\beta$) can always be set to a real parameter, which is equivalent to defining the relevant phases of two travelling waves by the imprinted modulation pattern. The last terms are source terms, and $P_{\circlearrowright}=|\tilde{S}_{\circlearrowright}|^2$ and $P_{\circlearrowleft}=|\tilde{S}_{\circlearrowleft}|^2$ represent the clockwise and counter-clockwise input powers in the waveguide. This interaction ends up renormalizing two propagating waves into two standing waves, $\tilde{A}_{\pm} = (\tilde{A}_{\circlearrowright}\pm\tilde{A}_{\circlearrowleft})/\sqrt{2}$, given by:
\begin{eqnarray}
\frac{d\tilde{A}_{\pm}}{dt} &=& [i(\Delta\omega \mp \beta) - \kappa_\text{0}/2]~\tilde{A}_\text{$\pm$} + i \sqrt{\kappa_\text{c}} \tilde{S}_{\pm}. \label{EqS3}
\end{eqnarray}
Here $\tilde{A}_{\pm}$ have the same cavity linewidth of $\kappa^0_\text{t} = \kappa_\text{0}+\kappa_\text{c}$, with the central resonance frequencies at $\omega_{\pm} = \omega_0 \pm \beta$, respectively. The source term follows a similar definition of $\tilde{S}_{\pm}=(\tilde{S}_{\circlearrowright}\pm\tilde{S}_{\circlearrowleft})/\sqrt{2}$.

From previous SMS works~\cite{lu2014selective, lu2020universal}, $\beta$ can be calculated for an optical mode with shifting boundaries~\cite{Johnson_PRE_2002},
\begin{eqnarray}
\beta = \frac{\omega_0}{2} \frac{\int{d S \cdot A \left[ (\epsilon_\text{d} - \epsilon_\text{c})|E_{\parallel}|^2 + (1/\epsilon_\text{c} - 1/\epsilon_\text{d})|D_{\perp}|^2\right]}}{\int d V {\epsilon(|E_{\parallel}|^2 + |E_{\perp}|^2)} }, \quad \label{EqS4}
\end{eqnarray}
where $E_{\parallel}$ ($D_{\parallel}$) and $E_{\perp}$ ($D_{\perp}$) are the electric field components (displacement field components) of the unperturbed optical mode that are parallel ($\parallel$) and perpendicular ($\perp$) to the modulation boundary $d S$, respectively. $\epsilon$ represents the dielectric function of the material, including the dielectric core ($\epsilon=\epsilon_\text{d}$) and the substrate ($\epsilon=\epsilon_\text{s}$) or cladding material ($\epsilon=\epsilon_\text{c}$). Considering a photonic crystal microring with a fixed outer radius but modulated ring width of $W(\phi) = W_\text{0} + \sum_{n}{A_n \text{cos}(n\phi)}$, the standing-wave mode with a larger frequency has a dominant displacement field $D(\rho,\phi,z)=D(\rho,z)\text{cos}(m \phi)$. Equation~(\ref{EqS4}) can be written as:
\begin{eqnarray}
\beta = \sum_{n} \frac{A_n \omega_0}{2} \frac{ \int{d S (1/\epsilon_\text{c} - 1/\epsilon_\text{d})|D(\rho,z)|^2 \text{cos}^2(m \phi) \text{cos}(n \phi) }}{\int {d V \epsilon (|E_{z}|^2+|E_{\phi}|^2+|E_{\rho}|^2) \text{cos}^2(m \phi) } }. \quad \label{EqS5}
\end{eqnarray}
This integral does not contain polar angle ($\theta$), assuming the dominant electric field is in the radial direction and the sidewall of the modulation is straight. The azimuthal part can be integrated separately, so that:
 \begin{eqnarray}
\beta = \sum_{n} g A_n \int_0^{2\pi}{d \phi \text{cos}^2(m \phi) \text{cos}(n \phi)} / \pi = \sum_{n} g A_n (\delta_{n,0}+\delta_{n,2m}/2) = g_{mm} (A_0 + \frac{A_{2m}}{2}), \label{EqS6}
\end{eqnarray}
where $\delta_{i,j}=1$ when $i = j$ and vanishes otherwise. $g$ is defined as:
\begin{eqnarray}
g \equiv \frac{\omega_0}{2} \frac{\int{d S (1/\epsilon_\text{c} - 1/\epsilon_\text{d})|D(\rho,z)|^2}}{\int {d V \epsilon (|E_{z}|^2+|E_{\phi}|^2+|E_{\rho}|^2) }  }. \quad \label{EqS7}
\end{eqnarray}

While we derive the Eqs.~(\ref{EqS4}-\ref{EqS7}) considering the shifting boundary for two standing-wave modes, this same parameter ($\beta$) is also the interaction term for two propagating waves described in Eqs.~(\ref{EqS1}-\ref{EqS2}). In particular, Eq.~(\ref{EqS6}) can be viewed as the interaction of clockwise ($m$) and counter-clockwise ($-m$) modes with the grating of ($n$) cells. The phase in the $\cos$ terms are set to zero here, which yields the maximal interaction/coupling rates, as the modes typically tend to maximize or minimize their energy. However, in specific cases the phase can be shifted, with one example being the phase shift for the bound standing-wave state at $l=0$ OAM, as discussed in the main text.

In the OAM cases, for the travelling wave modes, the coupled mode equations are given by:
\begin{eqnarray}
\frac{d\tilde{A}_{\circlearrowright}}{dt} &=& [i\Delta\omega - (\kappa_\text{0}/2+\beta')]~\tilde{A}_{\circlearrowright} + i \sqrt{\kappa_\text{c}} \tilde{S}_{\circlearrowright},  \label{EqS8} \\
\frac{d\tilde{A}_{\circlearrowleft}}{dt} &=& [i\Delta\omega - (\kappa_\text{0}/2+\beta')]~\tilde{A}_{\circlearrowleft} + i \sqrt{\kappa_\text{c}} \tilde{S}_{\circlearrowleft}. \label{EqS9}
\end{eqnarray}
There are no coupling of the clockwise and counter-clockwise modes induced by the grating. Instead, the grating introduces couplings from the WGM of azimuthal order $m$ to an ejected free-space OAM mode $l = m - N$ for $m < N < 2 m$ (and $-m$ to $-l$), and $\beta'$ is the scattering rate. For the standing-wave modes, either by accidental mode splitting created by the random sidewall roughness~\cite{Weiss_OL_1995}, or another added grating for SMS, as demonstrated in the last section in the main text, the equation is given by:
\begin{eqnarray}
\frac{d\tilde{A}_{\pm}}{dt} &=& [i\Delta\omega - (\kappa_\text{0}/2+\beta')]~\tilde{A}_\text{$\pm$} + i \sqrt{\kappa_\text{c}} \tilde{S}_{\pm}. \label{EqS10}
\end{eqnarray}
Here the $\beta'$ term represents the scattering rate from a standing wave mode with $|m|$ to OAM modes with $|l|$ (consisting of $\pm l$ equally).
\begin{figure*}[t!]
\centering\includegraphics[width=0.9\linewidth]{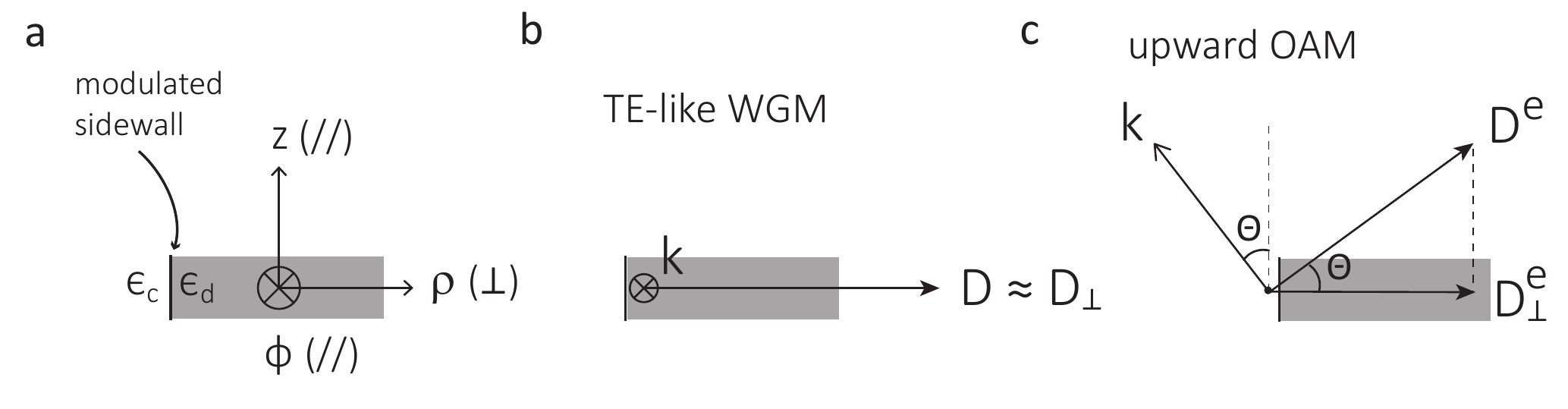}
\caption{Schematic illustration of the WGM and OAM fields at the modulated sidewall. \textbf{a} Cross-sectional drawing of the microring and the coordinates ($\rho$, $\phi$, $z$) in use. The device layer and the cladding have dielectric constants of $\epsilon_d$ (in gray) and $\epsilon_c$ (in white), respectively. The modulated sidewall is emphasized by a black solid line. \textbf{b} For a transverse-electric-like (TE-like) WGM, the dominant electric displacement is in the radial direction, that is, perpendicular to the sidewall in modulation. \textbf{c} For the ejected OAM mode, the dominant field direction depends on the nominal ejection angle ($\theta$) that is determined by angular momentum conservation. We illustrate upward OAM here, and downward OAM possesses mirror symmetry for this illustration.}
\label{SIFig1}
\end{figure*}

From Johnson et al.'s theory~\cite{Johnson_PRE_2002}, the calculation of the coupling of two modes is identical to the energy shift to one mode itself. The SMS is a perfect example in that $\beta$ can be considered as either the interaction of two travelling-wave modes, or the energy shift of a standing-wave mode from the modulated boundary. The OAM case is similar but more subtle: one can consider the OAM and WGM as being two modes, as the OAM is a radiating mode and the WGM is a confined mode. Yet, it is also not wrong to consider OAM and WGM to be two parts of the same mode in a broader definition, for the reason that the OAM mode is directly ejected out of the WGM mode, and they have the same mode profile before ejection (identical in polarization, linewidth, coupling, etc.). In both cases, if we follow Johnson et al.'s theory in a similar fashion,  we have $\beta'$ in the following:
\begin{eqnarray}
\beta' = \frac{\omega_0}{2} \frac{\int{d S \cdot A (1/\epsilon_\text{c} - 1/\epsilon_\text{d}) D_{\perp} D^e_{\perp}}}{\int d V {\epsilon(|E_{\parallel}|^2 + |E_{\perp}|^2)} }, \quad \label{EqS11}
\end{eqnarray}

The only difference between Eq.~\ref{EqS11} and Eq.~\ref{EqS4} is $D^e_{\perp}$ instead of $D_{\perp}$ (in both cases we assume that $E_{\parallel}$ is much smaller than the perpendicular field components). This term can be estimated from $D_{\perp}$ by considering three factors: (1) considering the ejected field has the same cross-sectional mode profile in the microring but without a cavity enhancement, we have $D^e=D/\sqrt{F_t/(2\pi)}$ since $F_t$ is defined in terms of energy (hence the square root) and phase (hence the $2\pi$); (2) since we are considering transverse-electric-like WGMs whose dominant field components are in the radial direction, we have $D_{\perp} \approx D$, as illustrated in SI Fig.~\ref{SIFig1}(a,b); and (3) the perpendicular projection can be estimated by the nominal ejection angle of the OAM beam from angular momentum conservation, as discussed in the main text, that is, $D^e_{\perp} = D^e \cos(\theta)$ and $\theta = (l/m)(\pi/2)$, as illustrated in SI Fig.~\ref{SIFig1}(c). Considering these three factors, Eq.~\ref{EqS11} can thus be written as
\begin{eqnarray}
\beta' = \sum_{n} \frac{A_n \omega_0}{2} \frac{ \int{d S (1/\epsilon_\text{c} - 1/\epsilon_\text{d})|D(r,z)|^2 \cos(m \phi) \cos(l \phi) \cos(n \phi) }}{\int {d V \epsilon (|E_{z}|^2+|E_{\phi}|^2+|E_{\rho}|^2) \text{cos}^2(m \phi)}} \frac{q_0\cos(\theta)}{\sqrt{F_t/(2\pi)}}. \quad \label{EqS12}
\end{eqnarray}
Here an additional term of $q_0$ is added in, considering that the emitted OAM light can be either upwards or downwards, which carries an additional factor of 2 compared to backward scattering in the SMS case. The value of this $q_0 = 2$ is validated in the experiments in the main text as a general trend, but its relationship to the light cone (and asymmetric cladding) requires further investigation. The azimuthal integral can be separated and yields similar results when $l = m-n$, so that Eq.~\ref{EqS12} can be reduced to:
 \begin{eqnarray}
\beta' =  \beta \frac{q_0 \cos(\theta)}{\sqrt{F_t/(2\pi)}}. \label{EqS13}
\end{eqnarray}
Considering the role of $\beta'$ in Eqs.~(\ref{EqS8}-\ref{EqS10}), we have Eq.~(1) in the main text,
 \begin{eqnarray}
\kappa_e = 2 \beta' = q_0  \frac{2\beta}{\sqrt{F_t/(2\pi)}} \cos(\theta). \label{EqS14}
\end{eqnarray}
Though we are considering standing-wave fields here as examples, the ejection rate ($\kappa_e$) is the same for travelling waves, similar to the coupling rate ($\beta$) that is used both in standing-wave and traveling-wave cases for SMS. Also, while our derivation is for transverse-electric-like fields, transverse-magnetic-like fields have the same results of Eq.~\ref{EqS14}, though the major contribution of $\beta$ and $\kappa_e$ is contributed from the parallel field components.

\section{Simulation Method}
\begin{figure*}[htbp!]
\centering\includegraphics[width=0.9\linewidth]{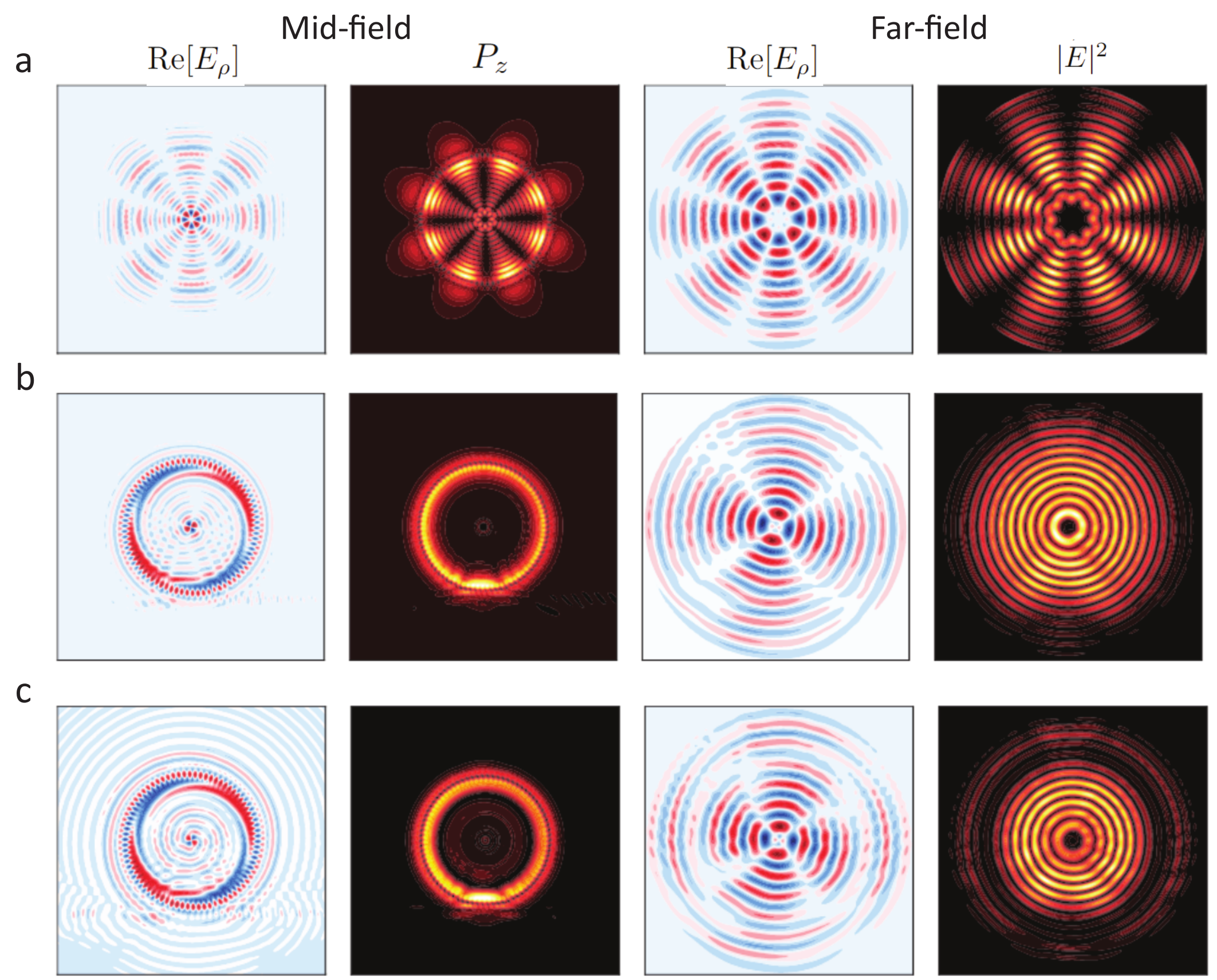}
\caption{Examples of cavity modes calculated by FDTD simulations. \textbf{a} Isolated ring with dipole excitation. \textbf{b} Bus waveguide-coupled ring with 3$\,\mu$m SiO$_2$ bottom cladding. \textbf{c} Bus waveguide-coupled ring wit 3$\,\mu$m thick SiO$_2$ bottom cladding and Si substrate. The four columns plot $|E|^2$, midfield: ${\rm{Re}}[E_\rho]$, $P_z$, far field: ${\rm{Re}}[E_\rho]$, $|E|^2$, respectively.} \label{SIFig2}
\end{figure*}

\noindent In this section, we discuss the method to simulate the mid-field and far-field images used in the main text. We employ finite-difference-time-domain (FDTD) simulations using the commercially available software Lumerical~\cite{NIST_disclaimer}. To reduce the required simulation resources, we consider smaller ring radii of 5~$\mu$m and 12$\,\mu$m as well as a modulation amplitude of $A=100\,$nm. We consider devices in three geometries. The first device, shown in SI Fig.~\ref{SIFig2}(a), consists of an isolated ring with air cladding on the top and bottom. The second device in SI Fig.~\ref{SIFig2}(b) has an additional bus waveguide below the ring and a 3~$\mu$m thick SiO$_2$ bottom cladding. The third device in SI Fig.~\ref{SIFig2}(c) has a bus waveguide below the ring, SiO$_2$ bottom cladding, and a Si substrate. The bus waveguide is 750~nm wide and separated from the ring by a gap of 350~nm. We use the refractive indices of $n_\text{SiN} = 2$,  $n_\text{SiO$_2$} = 1.45$, and $n_\text{Si} = 3.48$. The grid-spacing used is 30~nm in all three directions (note that we use Lumerical's ability to automatically set a coarser grid away from the ring).

SI Figure~\ref{SIFig2}(a) shows an example of the isolated ring with $N = 78$, which results in a resonant mode at $\lambda = 1677\,$nm with $m=74$, corresponding to $l=-4$. The modes are excited using 10 dipole sources evenly distributed over one period of the waveguide modulation inside the ring. Several dipoles are used to ensure that all longitudinal modes are excited. The mid-field profile plotted in SI Fig.~\ref{SIFig2}(a) is calculated as a running Fourier transform of the time-dependent field amplitude, $\mathbf{E}(t)$, in the plane at $z=0$. Note that Lumerical offers a setting to multiply the field amplitude by a ramping-function that eliminates any contribution from the sources. The dipole sources excite both clockwise and counterclockwise modes and the resulting mode shows an interference pattern between the two. The number of anti-nodes in $|\mathbf{E}|^2$ is therefore $2|l|$. In the mid-field images, ${\rm{Re}}[E_\rho]$ and $P_z$ are plotted in a plane 1$\,\mu$m above the waveguide surface, and in the far-field images, ${\rm{Re}}[E_{\rho}]$ and $|E|^2$ of the far field are calculated using Lumerical's built-in far field transformation routine. It is the field recorded 1~$\mu$m above the waveguide surface that is used in the transformation. Besides dipole excitation, standing-wave WGMs can be created by SMS (as described in the main text), random sidewall backscattering from the microring, and the chip facet reflection. The latter two typically lead to partial reflection instead of an equal contribution of clockwise and counter-clockwise modes, and will lead to more blurry (i.e., less visible) interference patterns. 

SI Figure~\ref{SIFig2}(b, c) shows mode profiles of structures including a bus waveguide with symmetric air cladding and with SiO$_2$ bottom cladding. SI Figure~\ref{SIFig2}(c) also has a Si substrate underneath the SiO$_2$ bottom cladding. For both structures, the source of the FDTD simulation is an eigenmode of the bus waveguide propagating from left to right and exciting only the counterclockwise ring mode. The mode profiles are again only monitored at times after the source has died out and the fact that only the counterclockwise mode is excited is verified by light only leaking out to the right in the bus waveguide. We can clearly see that, with such travelling-wave excitation, the field does not show $2|l|$ beating patterns in either mid-field $P_z$ or far-field $|E|^2$. In other words, the phase is varied by $l = -4$, but in intensity it is uniform.

In particular, SI Fig.~\ref{SIFig2}(c) suggests that the substrate reflection, that is, at the SiO$_2$/Si interface, cannot be used to explain the observed 2$|l|$ beating pattern, in contrast to the facet reflection as shown in SI Fig.~\ref{SIFig2}(a). Because of the symmetry, the device in simulation has $l = -4$ for the upward OAM (into air) and $l = 4$ for the downward OAM (into SiO$_2$). The $l = 4$ light is reflected at the SiO$_2$/Si interface to $l = -4$. Such reflected light with $l = -4$ can shift the phase of the original $l = -4$ light, but cannot create a beating pattern in intensity.

\section{Extracting $\kappa$ and $\beta$ from transmission spectra.} 
\begin{figure*}[htbp!]
\centering\includegraphics[width=0.95\linewidth]{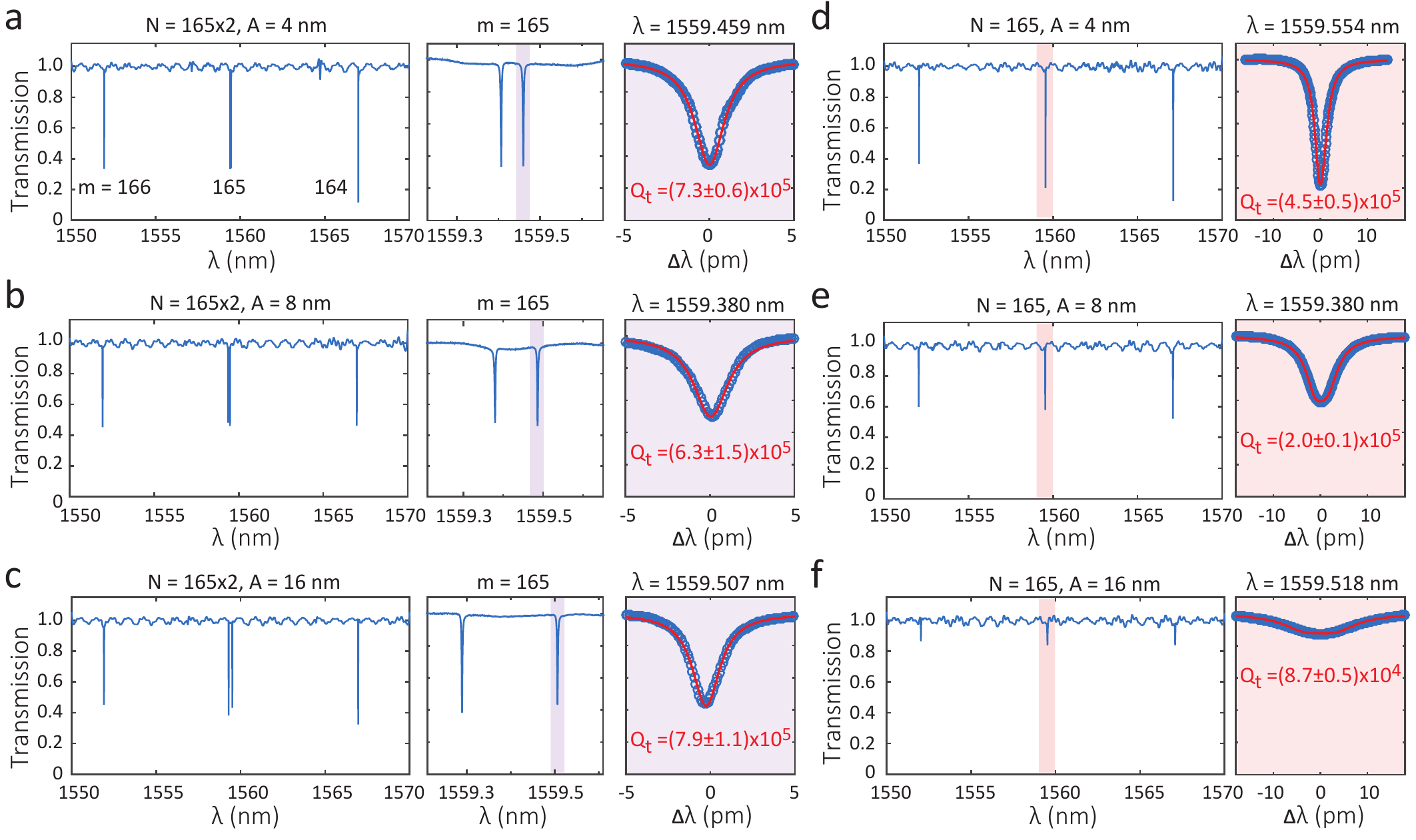}
\caption{Examples of characterization of selective mode splitting (SMS) and orbital angular momentum (OAM) devices. \textbf{a-c} Cavity transmission of the SMS devices with $N$ = 165$\times$2 and $A$ from 4~nm to 16~nm. The splitting increases when $A$ increases. \textbf{d-f} Cavity transmission of the OAM devices with $N$ = 165 and $A$ from 4~nm to 16~nm. The linewidth increases when $A$ increases. The uncertainty in $Q_t$ is given by a 95~\% confidence range of nonlinear fitting.}
\label{SIFig3}
\end{figure*}

\noindent We show here examples of extracting $\kappa$ and $\beta$ from representative transmission spectra for selective mode splitting (SMS) and orbital angular momentum (OAM) devices, more specifically, six devices with $l$ = $\{$-165, -60$\}$ and $A$ = $\{$4, 8, 16$\}$~nm in Fig.~3(c) in the main text. In SI Fig.~\ref{SIFig3}(a-c), SMS devices show increased mode splitting in the targeted mode when $A$ increases, but the overall optical quality factor ($Q_t$) is not strongly affected (i.e., remains within a common range of $Q_t$ values determined by the uncertainties in the nonlinear least squares fits to the transmission data). In SI Fig.~\ref{SIFig3}(d-f), OAM devices have decreased $Q_t$ and broader linewidths ($\kappa_t=\omega/Q_t$, which is used throughout the main text) when $A$ increases. The cavity transmission becomes under-coupled (shallow in dip) from the perspective of waveguide coupling, as the microring waveguide coupling rate ($\kappa_c$) stays the same while the OAM ejection rate ($\kappa_e$) increases.

\section{Comparing OAM devices with and without SMS.}
\begin{figure*}[htbp!]
\centering\includegraphics[width=0.9\linewidth]{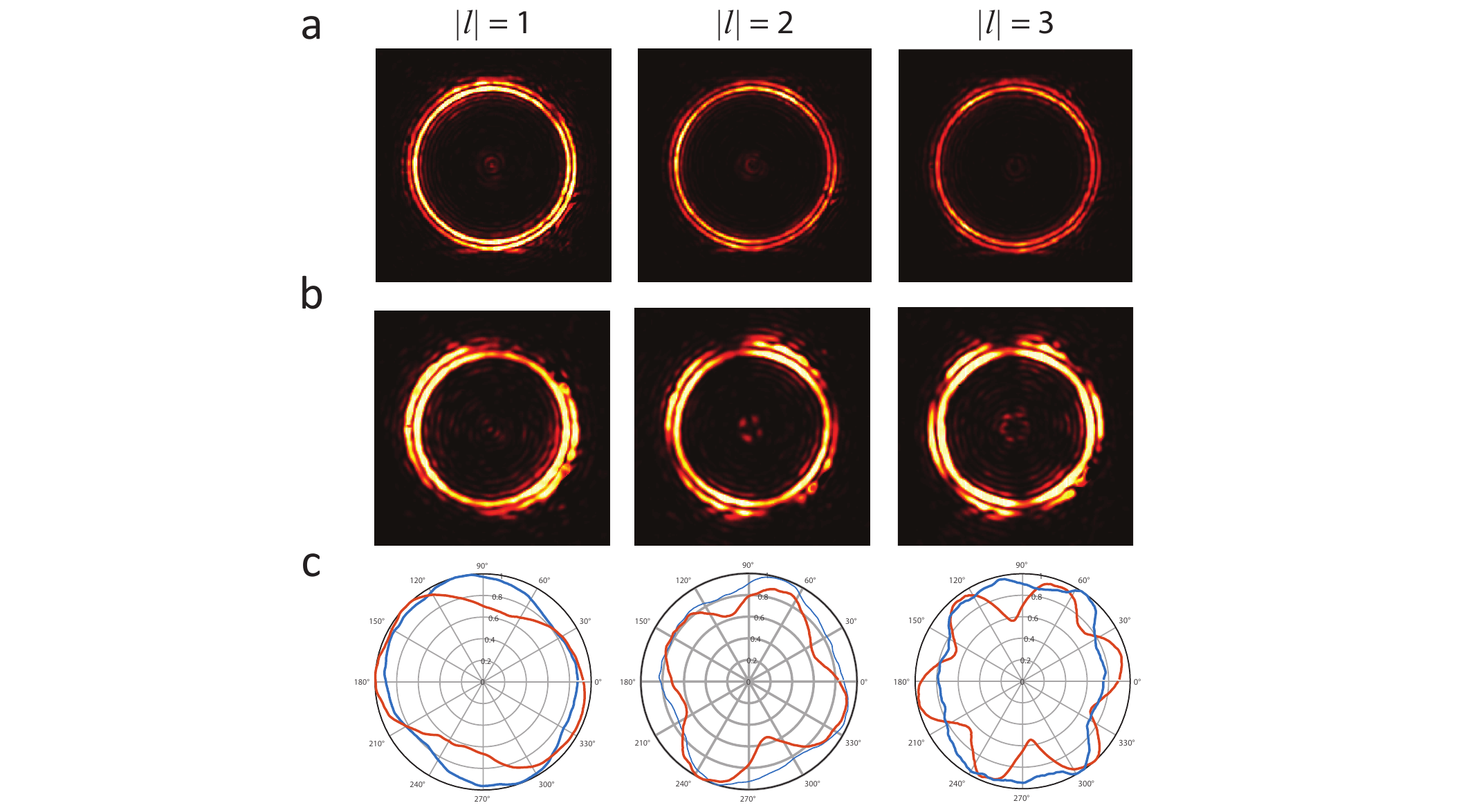}
\caption{Infrared images for the devices without and with SMS. \textbf{a} Device without SMS does not show clear visibility of the beating patterns. Here the mode has $m = 162$, and three devices have $N = \{$163, 164, 165$\}$, respectively. \textbf{b} Device with SMS showing clear beating patterns. Besides the OAM modulation in (a), an additional modulation with $N = 2 \times 162$ is added to couple and split the traveling wave modes into standing-wave modes. \textbf{c} Normalized intensity distribution as a function of azimuthal angle ($\phi$) for the devices without (blue) and with (red) SMS. The devices with SMS have a better visibility in the 2$|l|$ beating pattern.}
\label{SIFig4}
\end{figure*}

\noindent We compare OAM devices with and without SMS for $|l|$ = 1 to 3 in SI Fig.~\ref{SIFig4} . The OAM device without SMS has worse visibility in intensity beating, as the counter-clockwise WGM is likely from the back reflection of the pump laser, either from the chip facet or from the microring, and is thus much smaller in amplitude than the clockwise WGM. The method of implementing SMS and OAM has been discussed in the main text. The OAM device with SMS has standing-wave WGMs (equal clockwise and counter-clockwise WGMs) and hence nearly equal $+l$ and $-l$ OAM, resulting in more pronounced interference patterns.

\bibliographystyle{naturemag}

\begin{thebibliography}{1}
\bibitem{NIST_disclaimer}
\bibinfo{note}{Certain commercial products or names are identified to foster
  understanding. Such identification does not constitute recommendation or
  endorsement by the National Institute of Standards and Technology, nor is it
  intended to imply that the products or names identified are necessarily the
  best available for the purpose.}
\end{thebibliography}

\end{document}